\documentclass[twoside,11pt]{article}
\usepackage{amssymb,amsmath}

\usepackage{graphicx, subfigure, wrapfig}
\usepackage{amsfonts}
\usepackage[english]{babel}

\RequirePackage{color}

\newtheorem{theorem}{Theorem}

\newtheorem{remark}[theorem]{Remark}

\numberwithin{theorem}{section}
\newenvironment{proof}[1][Proof]{\textbf{#1.} }{\ \rule{0.5em}{0.5em}}

\DeclareMathOperator*{\trace}{tr}

\def\0{{\bf 0}}

\def\c{{\bf c}}
\def\d{{\mathrm{d}}}

\def\r{{\bf r}}

\def\p{{\bf p}}

\def\R{\mathbf{R}}
\def\P{\mathbf{P}}

\def\Gama{\boldsymbol{\Gamma}}
\def\Lam{\boldsymbol{\Lambda}}
\def\Om{\boldsymbol{\Omega}}
\def\Pii{\boldsymbol{\Pi}}
\def\u{{\bf u}}

\def\x{{\bf x}}

\def \beq{\begin{equation}}
\def \eeq{\end{equation}}
\def \intR {\int_{a}^{b}}
\def \lr  {\left\langle}
\def \rr   {\right\rangle}
\def \pl {\left (}

\def \d {\delta}

\begin{document}

\title{A note on the geometric modeling of the  full two body problem\\
Tanya Schmah\footnote{University of Ottawa, Email: tschmah@uottawa.ca}
\,\, and \,\, Cristina Stoica\footnote{Wilfrid Laurier University, Waterloo, Email: cstoica@wlu.ca}
}

\maketitle

\begin{abstract}
The two full body problem concerns the dynamics of two spatially extended rigid bodies (e.g. rocky asteroids) subject to mutual gravitational interaction.  In this note we deduce the  Euler-Poincar\'e  and Hamiltonian equations of motion using the geometric mechanics formalism.
\end{abstract}


\tableofcontents

\section{Introduction} 

It is well known that the classical two body problem, in which the bodies are idealized as mass points, can be analysed with almost elementary methods. 
Once the ``mass-point'' assumption is dropped, one is faced with a significantly more complex problem: a coupled, nonlinear  $12$ degrees of freedom system with a configuration space given by the product of two $SO(3)$ Lie  groups and two copies of $\mathbb{R}^3.$ The main inconvenience in modeling  resides in the lack of a global chart for $SO(3);$ 
for this reason, even for a single rigid body, most classical mechanics textbooks use Euler angles or alike, leading to an intricate  presentation; see for example, \cite{iacob1980theoretical}.


Anticipating  future developments in the aerospace industry, the full two body problem was studied extensively in the last decades; see for instance, \cite{Mac95},  \cite{Ko04},   \cite{Sc06},  \cite{BS08},  \cite{Sc09},  \cite{HX18} and references within. The modeling of the problem within the geometric mechanics  framework is developed in \cite{CM04}. However, this  presentation uses extensively   the geometric formalism  at an  abstract level. In this  note we provide a description of the full two body problem  within the geometric mechanics framework working \textit{directly} in the full two body problem phase space, and thus avoiding  abstract generalizations.

We start our modeling by assuming that the reduction  due to the linear translation symmetry has already been performed and that the centre of mass coincides with the origin of the  inertial system of coordinates. We write the Lagrangian, observe the $SO(3)$  symmetry and state and prove the appropriate (Euler-Poincar\'e) reduction theorem. We continue 
by computing the Euler equations.  Next, we  apply the reduced Legendre transform and deduce the  Poisson structure of the reduced space, the Hamiltonian, and the  equations of motion. Finally, we deduce the Casimir invariant as a consequence of the conservation of the size of the spatial angular momentum. We also include a small appendix with some formulae concerning the potential.

\section{Modeling and equations of motion}

Consider two  rigid  bodies 
moving freely in space, with a  coupling (gravitational) potential $V$ depending
on the orientations of the bodies and the relative position $\r$ of their centres of mass.
Choose a spatial coordinate system with origin at the centre of mass of the entire system,
which we assume remains fixed.
Let $\mathbf{r}_i$ be the vector from the centre of mass of the system to the centre of mass 
of body $i$, for each $i$. Let $\r = \mathbf{r}_2 - \mathbf{r}_1$.
Let $\mathcal{B}_1$ and $\mathcal{B}_2$, both subsets of $\mathbb{R}^3$,
be the reference configurations of the two rigid bodies,
each equipped with a reference frame defining \textit{body coordinates}, 
with origin at the body's centre of mass. 
A configuration of the system is determined by 
$\left(R_1, R_2, \mathbf{r}\right)$,
where $R_i$ specifies a rotation of body $i$ from its reference configuration,
around its own centre of mass  (see, for instance, \cite{MaRa1999}).
The configuration space of the system is $Q := SO(3) \times SO(3) \times \mathbb{R}^3 \setminus\{\text{collisions}\}$, where $ SO(3)$ denotes the Lie group of spatial rotations.

Let $\mu_i$ be the mass measure for body $i$, for $i=1,2$.  Then the total
mass of body $i$ is
\[
m_i:= \int_{\mathcal{B}_i} d\mu_i.
\]
The translational kinetic energy of body $i$ is 
$\frac{1}{2} m_i \left\| \dot {\mathbf{r}}_i\right\|^2$. Following the centre of mass  reduction, the \emph{reduced mass} is $m := \frac{m_1m_2}{m_1 + m_2}$ and the total translational kinetic energy of the system is
$\frac{1}{2} m \left\| \dot {\mathbf{r}}\right\|^2$.

The \emph{coefficient of inertia matrix} of body $i$, with respect to its own centre of mass, is
\begin{align*}
\mathbb{J}_i := \int_{\mathcal{B}_i} XX^t \, d\mu_i (X),
\end{align*}
where $({\cdot})^t$ denotes the matrix transpose. 
The body angular velocities are
$
\hat \Omega_i := {R_i^{-1} \dot R_i}.
$
The rotational kinetic energy of body $i$ is
\begin{align*}
K_i=\frac{1}{2} \lr  \dot R_i\,,\dot R_i   \rr_i:=\frac{1}{2} \trace \left(\dot R_i {\mathbb{J}}_i \dot R_i^t\right) 
=\frac{1}{2} \trace \left(\left({R_i^{-1} \dot R_i}\right) {\mathbb{J}}_i  \left({R_i^{-1} \dot R_i}\right)^t\right) 
=\frac{1}{2} \trace \left(\hat \Omega_i {\mathbb{J}}_i \hat \Omega_i^t\right) 
= :\frac{1}{2} \lr\hat \Omega_i\,,\hat \Omega_i \rr_i  \, .
\end{align*}
The \emph{moment of inertia tensors} are
\begin{equation}
{\mathbb{I}}_i := \trace\left({\mathbb{J}}_i\right) \text{Id}_3 - {\mathbb{J}}_i\, ,
\label{in_tens}
\end{equation}
where $\text{Id}_3$ is the $3 \times 3$ identity matrix.
Using  the usual identification of  the Lie algebra $so(3)$ with $\mathbb{R}^3$ via the hat map  ${\,\,\,\,\,\hat{}}: \mathbb{R}^3 \to so(3),$ 
 \[
\Omega=(\Omega_1,\Omega_2,\Omega_3)\rightarrow \hat \Omega=\left[\begin{array}{ccc}
0&-\Omega_3& \Omega_2\\
\Omega_3&0&-\Omega_1\\
-\Omega_2&\Omega_1& 0\\
\end{array}\right],
\] 
we can also write 
\[
K_i =\frac{1}{2} \trace \left(\hat \Omega_i {\mathbb{J}}_i \hat \Omega_i^t\right)  = \frac{1}{2} \Omega_i^t {\mathbb{I}}_i \Omega_i \,.
\]
For further reference, recall that for any matrices $\hat  \Omega\,, \hat \Lambda \in so(3)$ corresponding to the vectors $  \Om\,,  \Lam \in \mathbb{R}^3$, we have
\[
\left[ \hat  \Omega\,, \hat \Lambda \right] =  \Om \times \Lam
\]
where $[\cdot\,, \cdot]$ denotes the matrix Lie-bracket (i.e. $[A,B] = AB-BA$)\,.

\bigskip

%
In coordinates on the tangent bundle $ T(SO(3) \times SO(3) \times \mathbb{R}^3\setminus\{\text{collisions}\})$,
 the dynamics is given by the Lagrangian
\begin{equation} 
L(R_1, R_2,\r, \dot R_1, \dot R_2, \dot \r)=\frac{1}{2} \lr  \dot R_1\,,\dot R_1   \rr_1+ \frac{1}{2} \lr  \dot R_2\,,\dot R_2  \rr_2\, +\frac{1}{2}m\|\dot \r\|^2-V(R_1,R_2, \r)\,.
\label{Lagrangian}
\end{equation}

\bigskip
\noindent
The spatial action of $SO(3)$ on the configuration space is the diagonal 
left multiplication action,
\begin{equation}
A \cdot \left(R_1, R_2, \r\right) = \left(A R_1, A R_2, A \r\right)\,,\quad \quad A \in SO(3).
\label{action}
\end{equation}
Since $L$ is invariant under this action, the dynamics may be retrieved from a reduced system. Indeed, describing   the motion in the coordinates of one of the bodies allows us to render the equations as a reduced system on a smaller dimensional phase space (the reduced space), together with the so-called reconstruction equation that lifts the reduced dynamics back into the unreduced phase space. 

For future reference, we note that the infinitesimal action of $so(3)$ to $SO(3) \times SO(3) \times \mathbb{R}^3$ is (see \cite{HSS09}):
\begin{equation}
\hat \Omega_{SO(3) \times SO(3) \times \mathbb{R}^3} \cdot (R_1, R_2, \r) = (\hat \Omega R_1, \hat \Omega R_2, \hat \Omega \r)
\label{inf_so(3)}
\end{equation}

Denote the relative orientation matrix of ${\cal B}_2$  with respect to body ${\cal B}_1$, and  the relative position of the centre of the mass of the system, respectively, by
\begin{equation} 
R:= R_1^{-1}R_2 \quad \quad \text{and} \quad \Gama:=R_1^{-1}\r\,.
\label{rel_R}
\end{equation}
We then calculate the tangent vector (velocity corresponding to the relative orientation) $\dot{R} \in T_R SO(3)$ and the advected  relative velocity (i.e. the velocity corresonding to the relative vector) $\dot \Gama$
\begin{equation} 
 \dot{R}= R \hat\Omega_2- \hat\Omega_1R \quad \quad  \text{and} \quad \dot \Gama= R_1^{-1} \dot \r - \hat \Omega_1\Gama.
  \label{dot-O-and-G}
\end{equation}

Recalling that $\dot R_i=R_i \hat \Omega_i,$ $i=1,2,$ and using the above we calculate 
\begin{align*} 
L(R_1,R_2,\r,\dot R_1, \dot R_2, \dot \r)&=L(R_1^{-1}R_1,R_1^{-1}R_2,R_1^{-1}\r,R_1^{-1}\dot R_1, R_1^{-1}\dot R_2, R_1^{-1}\dot \r ) \\
&= L\left(R_1^{-1}R_1,R_1^{-1}R_2,R_1^{-1} (R_1 \Gamma), R_1^{-1}(R_1 \hat \Omega_1), R_1^{-1}(R_2 \hat \Omega_2), R_1^{-1}R_1( \dot\Gama +\hat\Omega_1\Gama )\right)\\
&=L(\text{Id}_3, R, \Gama, \hat\Omega_1, R\hat\Omega_2,\dot\Gama +\hat\Omega_1\Gama)
\end{align*}
from where we define the \textit{reduced lagrangian}
\begin{align} 
& l:SO(3)\times so(3)\times so(3)\times T \left({\mathbb R}^3 \setminus \{\text{collisions}\}\right)\rightarrow \mathbb{R} \nonumber \\ 
 &l(R,\hat\Omega_1,\hat\Omega_2,\Gama, \dot\Gama):=L(\text{Id}_3, R, \Gama, \hat\Omega_1, R\hat\Omega_2,\dot\Gama +\hat\Omega_1\Gama)
\end{align}
that takes the form 
\beq 
l(R,\hat\Omega_1,\hat\Omega_2,\Gama, \dot\Gama)=\frac{1}{2} \lr \hat\Omega_1\,,  \hat\Omega_1\rr_1 +\frac{1}{2} \lr \hat\Omega_2\,, \hat\Omega_2\rr_2 +\frac{1}{2} m\|  \dot\Gama +\hat\Omega_1\Gama \|^2-V(R,\Gama).
 \label{red-lag}
\eeq

Let $\left< \cdot \,,\cdot \right>_{\mathbb{R}^3}$ be the usual dot product on $\mathbb{R}^3.$  Thus,   for all $\boldsymbol{\Pi}  =  (\Pi_1\,,\Pi_2 \,,\Pi_3)\in \mathbb{R}^3  \simeq so(3)^*$ and $\boldsymbol{\Omega} = (\Omega_1,\Omega_2,\Omega_3)\in \mathbb{R}^3  \simeq so(3)$ 
we have 
\[
\left< \boldsymbol{\Pi},\, \boldsymbol{\Omega} \right>_{\mathbb{R}^3} =  \boldsymbol{\Pi} \cdot  \boldsymbol{\Omega}  =  \Pi_1 \,  \Omega_1 +\Pi_2   \,\Omega_2+\Pi_3 \,  \Omega_3\,.
\]
We denote the  pairing  between  $so(3)^*$ and  $so(3)$ in matrix notation by $\left< \cdot \,,  \cdot  \right>$ (no subscript!), and define the `breve' map, $\, \breve{} \, : \R^3 \rightarrow so(3)^*$, by
$\left< \breve\Pi\,, \, \hat \Omega \right> = \left< \Pi \,,\Omega \right>_{\mathbb{R}^3}$.
%
%
It can be shown that
\[
\left< \breve\Pi\,, \, \hat \Omega \right>=\frac{1}{2}\text{tr} (\hat \Pi^t  \, \hat \Omega)  =  \frac{1}{2}\text{tr} (\hat \Pi  \, \hat \Omega^t) 
\]
for all $\breve \Pi \in so(3)^*$ and $\Omega \in so(3)$.
%
%
%
%
\begin{equation}
\left<\hat M, R\hat \Omega \right> =  \text{tr}\left(\hat M^t (R\hat \Omega) \right) =  
 \text{tr}\left( (R^t \hat M)^t  \hat \Omega  \right)  =  \left<R^t \hat M \,,\hat \Omega \right>  =
  \left<(R^t \hat M)_A\,,\hat \Omega \right>   
  = \left< \left(R^t \hat M - \hat M^t R \right), \,\hat \Omega \right> 
  \label{left_star}
\end{equation}
for all $\hat \Omega \in so(3)$, where a matrix subscript $\,_A$ denotes the anti-symmetric part of that matrix. (We use here the fact that the trace pairing of any symmetric matrix with an antisymmetric matrix vanishes.) Similarly,  
\begin{equation}
\left<\hat M, \hat \Omega R \right> =  \text{tr}\left(\hat M^t (\hat \Omega R) \right) =  
 \text{tr}\left( (\hat M R^t )^t  \hat \Omega  \right)  =  \left< \hat M R^t \,,\hat \Omega \right>  =
  \left<(\hat M R^t)_A\,,\hat \Omega \right>   
  = \left< \left( \hat M R^t- R \hat M^t \right),\, \hat \Omega \right>. 
  \label{right_star}
\end{equation}

We are ready now to state the main theorem.

\bigskip

\begin{theorem} 
\label{EL_full_two}
Consider a Lagrangian  $L: T\left(SO(3) \times SO(3) \times D\right) \to \mathbb{R}$, $D \subset \mathbb{R}^3$ open, 
\[
L=L\left(R_1,R_2,\r,\dot R_1,\dot R_2, \dot \r  \right).
\]
For any  given curves $(R_1(t), R_2(t)) \in SO(3) \times SO(3)$ and $\r(t) \in \mathbb{R}^3$, let $R(t)= R_1^{-1}(t)R_2(t)$,  $\Gama(t) = R_1(t) \r(t)$ and 
\[
\hat \Omega_i (t):= R_i(t)^{-1}\dot R_i(t) \in so(3)\,.
\]
Consider 
\[
l(R,\hat\Omega_1,\hat\Omega_2,\Gama, \dot\Gama):=L(\text{Id}_3, R, \Gama, \hat\Omega_1, R\hat\Omega_2,\dot\Gama +\hat\Omega_1\Gama)
\]
and let   $R(t)$  be the solution of the non-autonomous differential equation
\begin{equation}
 \label{R-dynamics} \dot{R}= R(t) \hat\Omega_2(t)- \hat\Omega_1(t)R(t)\,,\quad R(0)= R_0\,.
 \end{equation}
where $R_0= R_1(0)^{-1} R_2(0)\,.$ The following statements are equivalent:

  (i)  $(R_1(t),R_2(t),\r(t))$ satisfies the Euler-Lagrange equations for the Lagrangian $L.$ 

(ii)  The  variational principle
\[
\delta \int_a^b L\left(R_1(t),R_2(t),\r(t),\dot R_1(t),\dot R_2(t), \dot \r(t)  \right)dt=0
\]
holds for variations with fixed endpoints.

  (iii)  The \textbf{reduced} variational principle 
\[
\delta \int_a^b l \left(R(t),\hat\Omega_1(t),\hat\Omega_2(t), \Gama(t), \dot\Gama(t) \right)dt=0
\]
holds using variations of the form  
\[\d\hat\Omega_i=\dot{\hat\Sigma}_i+ [\hat\Omega_i,  \hat\Sigma_i ] \quad \text{and} \quad \d \Gama=\Lam -\hat\Sigma_1\Gama  
\]
where the $\hat\Sigma_i(t)$ are arbitrary paths in $so(3)$ which vanish at the endpoints, i.e. $\hat\Sigma_i(a)=\hat\Sigma_i(b)=\hat 0$, $i=1,2$, and $\Lam(t)$ is an arbitrary path in $\mathbb{R}^3$ with $\Lam(a)=\Lam(b)={\bf 0}_{\mathbb{R}^3}.$

(iv) The (left invariant) ``Euler-Poincar\'e" equations hold: 
\begin{align}
&\frac{d}{dt}\left( \frac{\d l}{\d \hat\Omega_1} \right)=\left[\frac{\d l}{\d \hat\Omega_1}, {\hat\Omega_1} \right]
+\left(  R \left( \frac{\d l}{\d R} \right)^t-   \frac{\d l}{\d R}  R^t \right),  \\
&\frac{d}{dt}\left( \frac{\d l}{\d \hat\Omega_2} \right)=\left[\frac{\d l}{\d \hat\Omega_2}, {\hat\Omega_2} \right]
+ \left( R^t \frac{\d l}{\d R} -  \left( \frac{\d l}{\d R} \right)^t R \right),  \\
&\frac{d}{dt}\left(\frac{\d l}{\d\dot\Gama} \right)=\frac{\d l}{\d \Gama}.
\label{ELequations}
\end{align}

\end{theorem}

\bigskip

\begin{proof}
The equivalence of (i) and (ii) is a restatement  of Hamilton's principle.
To show that (ii) and (iii) are equivalent, we compute the variations $\delta \hat\Omega_1,\d\hat\Omega_2,$ and $\d\Gama$ and induced by the variations $\d R_1, \d R_2,$ and  $\d \r.$

Given that $\hat\Omega_i=R_i^{-1}\dot{R}_i$ and denoting $\hat\Sigma_i:= R_i^{-1}\delta R_i \in so(3),$ $i=1,2$ we calculate: 
\[\begin{split}
\delta \hat\Omega_i &=(\delta R_i^{-1}) \dot{R_i} +R_i^{-1} \delta\dot{R}_i= -(R_i^{-1} \delta{R}_i R_i^{-1})\dot{R}_i + R_i^{-1} \delta\dot{R}_i\\
&= -(R_i^{-1} \delta{R}_i)( R_i^{-1}\dot{R}_i)+R_i^{-1}\left(\delta\frac{dR_i}{dt}\right)
= -\hat\Sigma \hat\Omega_i +R_i^{-1}\frac{d}{dt}(\delta R_i)\\&= -\hat\Sigma_i \hat\Omega_i+\frac{d}{dt} \left( R_i^{-1}\delta R_i \right) - \dot{R_i}^{-1}\delta R_i= -\hat\Sigma_i \hat\Omega_i+\frac{d\hat\Sigma_i}{dt}+(R_i^{-1} \dot{R}_iR_i^{-1})\delta R_i\\ &= -\hat\Sigma_i \hat\Omega_i+\frac{d\hat\Sigma_i}{dt}+(R_i^{-1} \dot{R}_i)(R_i^{-1})\delta R_i=\frac{d\hat\Sigma_i}{dt}-\hat\Sigma_i \hat\Omega_i+ \hat\Omega_i \hat\Sigma_i\\ &= \frac{d\hat\Sigma_i}{dt}+[\hat\Sigma_i, \hat\Omega_i].
\end{split}\] 
Thus we have  
\[\d\hat\Omega_i=\frac{d\hat\Sigma_i}{dt}+[\hat\Sigma_i, \hat\Omega_i], \quad
i=1,2.
\]
The variation of $\Gama$ is
\begin{equation}  
\d\Gama=\d(R_1^{-1} \r)=\d(R_1^{-1}) \r+R_1^{-1}\d \r=-R_1^{-1}  (\d R_1)R_1^{-1} \r+R_1^{-1}\d \r
\end{equation}
Denoting $\Lam:=R_1^{-1}\d \r,$  the above reads:
\begin{equation}  
\d \Gama=\Lam-\hat\Sigma_1\Gama.
\label{d_gama}
\end{equation}
To complete the proof we show the equivalence of (iii) and (iv).
First note that  since  
\[\begin{split}
\d R&=\d(R_1^{-1}R_2)=\d(R_1^{-1})R_2+R_1^{-1}\d r_2=-(R_1^{-1}(\d R_1)R_1^{-1})R_2+R_1^{-1}R_2R_2^{-1}\d R_2\\
&=-(R_1^{-1}\d R_1)(R_1^{-1}R_2)+(R_1^{-1}R_2)(R_2^{-1}\d R_2)=-\hat\Sigma_1 R+R\hat\Sigma_2
\end{split}.\]
we have
\[
\d R=R\hat\Sigma_2-\hat\Sigma_1 R.
\]
Now we calculate
\begin{align}
& \d \intR l \left(R,\Gama,\hat\Omega_1,\hat\Omega_2,\dot\Gama \right)dt 
= \intR \left<  \frac{\d l}{\d R},\d R \right>
+ \left<  \frac{\d l}{\d \Gama},\d\Gama \right>_{\mathbb{R}^3}
+\sum_{i=1}^2\left<   \frac{\d l}{\d \hat\Omega_i},\d \hat\Omega_i  \right>
+\left<   \frac{\d l}{\d \dot \Gama},\d \dot \Gama \right>_{\mathbb{R}^3}   dt  \nonumber \\  
=&\intR \left< \frac{\d l}{\d R},R\hat\Sigma_2-\hat\Sigma_1 R \right>dt   
+ \left<  \frac{\d l}{\d \Gama},\d\Gama \right>_{\mathbb{R}^3}+
 \sum_{i=1}^2 \intR  \left<  \frac{\d l}{\d \hat\Omega_i}, \dot {\hat\Sigma}_i +[\hat\Sigma_i, \hat\Omega_i]   \right>dt 
+\left<   \frac{\d l}{\d \dot \Gama},\d \dot \Gama \right>_{\mathbb{R}^3}   dt
\label{calc_1}
\end{align}
Using the relations \eqref{left_star} and \eqref{right_star}, the first term of \eqref{calc_1} becomes
\begin{align*}
\intR \left< \frac{\d l}{\d R},R\hat\Sigma_2-\hat\Sigma_1 R \right>dt
&= 
\intR \left< \frac{\d l}{\d R},R\hat\Sigma_2  \right>dt  - \intR \left< \frac{\d l}{\d R},\hat\Sigma_1 R \right>dt\\
&=
\intR \left<\left( R^t \frac{\d l}{\d R} -  \left( \frac{\d l}{\d R} \right)^t R \right), \hat\Sigma_2  \right>dt  
- \intR
 \left< \left(     \frac{\d l}{\d R}  R^t -R \left( \frac{\d l}{\d R} \right)^t\right)
,\hat\Sigma_1  \right>dt
\end{align*}
Using that $\hat \Pi \in so^*(3)$ we have $ \left< \hat \Pi, [\hat\Sigma, \hat\Omega]  \right> =  \left<  [ \hat \Omega,  \hat \Pi],   \hat\Sigma   \right> $ for  all $\hat\Sigma, \hat\Omega \in so(3).$ 
 that $\d (d/dt) =(d/dt) \d,$  integrating by parts and taking into account the boundary conditions, the third term of  \eqref{calc_1} becomes:
\[ \intR \left< -\frac{d}{dt}\left( \frac{\d l}{\d \hat\Omega_1} \right)+ \left[\frac{\d l}{\d \hat\Omega_1}, \,{\hat\Omega_1} \right], \hat\Sigma_1   \right> dt
+ \intR \left< -\frac{d}{dt}\left( \frac{\d l}{\d \hat\Omega_2} \right)+ \left[\frac{\d l}{\d \hat\Omega_2}, {\hat\Omega_2} \right], \hat\Sigma_2   \right>  dt\]
 Finally,  define $\Gama \diamond \P \in so^*(3)$ via 
 $\left< \Gama \diamond \P ,  \hat\Sigma   \right>  := \left<\P, \hat \Sigma \Gama  \rr =\left<  \P,  \boldsymbol{\Sigma}  \times  \Gama   \right>_{\mathbb{R}^3} =\left<  \Gama \times \P ,  \boldsymbol{\Sigma}  \right>_{\mathbb{R}^3} $
 for all $\P, \Gama \in \mathbb{R}^3,$ and $\hat\Sigma \in so(3).$ 
 Substituting \eqref{d_gama}  the second and the last terms of \eqref{calc_1} transform to
\begin{align*}  \intR  \left< \frac{\d l}{\d \Gama}, \d\Gama  \right>_{\mathbb{R}^3} 
+ \left< \frac{\d l}{\d \dot\Gama},  \frac{d}{dt}\d  \Gama \right>_{\mathbb{R}^3}   
&= \intR  \left<   -  \frac{d}{dt}  \left(  \frac{\d l}{\d \dot\Gama}\right)+  \frac{\d l}{\d \Gama},  \Lam -  \hat \Sigma_1 \Gama \right>_{\mathbb{R}^3}  \\
 = & \intR  \left<  -  \frac{d}{dt}  \left(  \frac{\d l}{\d \dot\Gama}\right)+  \frac{\d l}{\d \Gama},  \Lam \right> _{\mathbb{R}^3} - \intR  \left<  \Gama \diamond \left[  -  \frac{d}{dt}  \pl \frac{\d l}{\d \dot\Gama}\right)+  \frac{\d l}{\d \Gama} \right],    \hat \Sigma_1\right> .
\end{align*}
Thus we obtain 
\begin{align*}  &\d \intR l \left(R,\Gama,\hat\Omega_1,\hat\Omega_2,\dot\Gama \right)dt \\
&\quad  =  \intR \left<  -\frac{d}{dt}\left( \frac{\d l}{\d \hat\Omega_1} \right)+ \left[\frac{\d l}{\d \hat\Omega_1}, {\hat\Omega_1} \right] -\left(     \frac{\d l}{\d R}  R^t -R \left( \frac{\d l}{\d R} \right)^t\right)-\Gama \diamond \left[  -  \frac{d}{dt}  \pl \frac{\d l}{\d \dot\Gama}\right)+  \frac{\d l}{\d \Gama} \right], \, \hat\Sigma_1  \right> dt \\ 
&\quad +  \intR \left<-\frac{d}{dt}\left( \frac{\d l}{\d \hat\Omega_2} \right)+ \left[\frac{\d l}{\d \hat\Omega_2}, {\hat\Omega_2} \right]+ \left( R^t \frac{\d l}{\d R} -  \left( \frac{\d l}{\d R} \right)^t R \right),\,\hat\Sigma_2  \right> dt   + \intR \left< -  \frac{d}{dt}  \lr  \frac{\d l}{\d \dot\Gama}\right)+  \frac{\d l}{\d \Gama},  \Lam \right>_{\mathbb{R}^3}dt.\end{align*}
Since $\hat\Sigma_1,\hat\Sigma_2$ and $\Lam$ are arbitrary, the conclusion follows.

\end{proof}

\bigskip  %
Recall that any  orthogonal matrix $R$ can be expressed as  $R:= [\boldsymbol{\alpha}_1, \boldsymbol{\alpha}_2, \boldsymbol{\alpha}_3]$ with $\boldsymbol{\alpha}_i \in \mathbf{R}^3,$ $i=1,2,3,$ such that $\boldsymbol{\alpha}_i^2=1$ and $\boldsymbol{\alpha}_i \cdot \boldsymbol{\alpha}_j=0$ for $i\neq j.$ Then for any function depending on $R\in SO(3)$, i.e., $f =f(R\,,\cdot) \to \mathbb{R}$  the vector representation of 
\begin{align*}
\hat T_1:= R \left( \frac{\d f}{\d R} \right)^t-   \frac{\d f}{\d R}  R^t   \quad &\text{and}\quad 
\hat T_2:= R^t \frac{\d f}{\d R} -  \left( \frac{\d f}{\d R} \right)^t R
\end{align*}
is
\begin{align*}
T_1= \sum \limits_{i=1,2,3} \boldsymbol{\alpha}_i \times \frac{\d f}{\d \boldsymbol{\alpha}_i} \quad &\text{and}
\quad T_2= -\sum \limits_{i=1,2,3} \boldsymbol{\alpha}_i \times \frac{\d f}{\d \boldsymbol{\alpha}_i}\,, \,\,
\end{align*}
respectively. Note that in the above, we calculate $\frac { \d f}{\d R}$ as the matrix
\[
\frac { \d f}{\d R}= 
\left[
\frac{\partial f }{\partial  \boldsymbol{\alpha}_1}\, \,\, \frac{\partial f }{\partial  \boldsymbol{\alpha}_2 }\, \,\,\frac{\partial f }{\partial  \boldsymbol{\alpha}_3} 
\right]
\]
where for the vector $\boldsymbol{\alpha}_i = (\alpha_{i1}\,,\alpha_{i2}\,,\alpha_{i3})^t$ we have $\displaystyle{\frac{\partial f }{\partial  \boldsymbol{\alpha}_i} = \left(\frac{\partial f }{\alpha_{i2} }\,,\,\, \frac{\partial f }{\alpha_{i1} }\,,\,\,\frac{\partial f }{\alpha_{i3} }  \right)^t }.$
This allows to writing the vector form of the reduced equations of motion (\ref{ELequations}):
\begin{align}
&\frac{d}{dt}\left( \frac{\d l}{\d \Om_1} \right)= \frac{\d l}{\d \Om_1} \times  \Om_1 
+\sum \limits_{i=1,2,3} \boldsymbol{\alpha}_i \times \frac{\d l}{\d \boldsymbol{\alpha}_i}  \\
&\frac{d}{dt}\left( \frac{\d l}{\d \Om_2} \right)=\frac{\d l}{\d \Om_2} \times \Om_2
-\sum \limits_{i=1,2,3} \boldsymbol{\alpha}_i \times \frac{\d l}{\d \boldsymbol{\alpha}_i} \\
&\frac{d}{dt}\left(\frac{\d l}{\d\dot\Gama} \right)=\frac{\d l}{\d \Gama}.
\label{ELeq-vector}
\end{align}
This above system is completed by the relative orientation equation (\ref{R-dynamics}).

Specializing the Lagrangian  to the full two body problem, the reduced lagrangian is given by \eqref{red-lag}. In vectorial notation the reduced lagrangian   is 
 \begin{equation}
l(R,\Om_1,\Om_2,\Gamma, \dot\Gamma):=\frac {1}{2} \left<\Om_1, \mathbb{I}_1 \Om_1\right>_{\mathbb{R}^3} +\frac {1}{2}\left< \Om_2, \mathbb{I}_2 \Om_2 \right>_{\mathbb{R}^3}+\frac {m}{2}\|  \dot\Gama + \Om_1 \times \Gama \|^2-V(R,\Gama)
  \label{red-l-vector}
 \end{equation}

and the equations of motion are 
\begin{align}
 \label{EL-eq-here}
&\frac{d}{dt} \left(  \mathbb{I}_1 \Om_1 + m \,\Gama \times   (\dot \Gama + \Om_1 \times \Gama)  \right)= \mathbb{I}_1 {\Om}_1 \times  {\Om}_1 
-  m \left[ (\dot \Gama + \Om_1 \times \Gama) \times \Gama \right]\times \Om_1
 +\sum \limits_{i=1,2,3} \boldsymbol{\alpha}_i \times \frac{\d V}{\d \boldsymbol{\alpha}_i}
 \\
&\mathbb{I}_2 \dot{\Om}_2 =\mathbb{I}_2\,{\Om}_2  \times {\Om_2} 
-\sum \limits_{i=1,2,3} \boldsymbol{\alpha}_i \times \frac{\d V}{\d \boldsymbol{\alpha}_i}
\\
&\frac{d}{dt}\left( \dot\Gama + \Om_1 \times \Gama \right)=\left( \dot\Gama + \Om_1 \times \Gama \right) \times  \Om_1  - \frac{1}{m}\frac{\d V}{\d \Gama}
\end{align}

\section{Hamiltonian formulation}

The Hamiltonian of the full two body problem may be obtained by applying the Legendre transform to the Lagrangian \eqref{Lagrangian} 
and it reads:
\begin{align}
&H: T^*SO(3) \times T^*SO(3) \times T^*\mathbb{R}^3\setminus\{\text{collisions}\} \to \mathbb{R} \nonumber \\
&H(R_1, \pi_{R_1}, R_2, \pi_{R_2}, \r,    \p)=\frac{1}{2} \left<\pi_{R_1}\,, \pi_{R_1}  \right>^*_1 + 
\frac{1}{2} \left< \pi_{R_2}\,, \pi_{R_2}  \right>^*_2+ \frac{1}{2m} \p^2 + V(R, \r)\,,
\label{Ham_2_B}
\end{align}
where the pairings $\left< \cdot\,,\cdot \right>^*_i$ on $T^*_{R_i}(SO(3)$ for fixed $R_i$, i=1,2 correspond to the kinetic terms in \eqref{Lagrangian}, and, as usual:
\begin{align}
\pi_{R_i} =\frac{\partial L}{\partial \dot R_i} \in T_{R_i}^*SO(3)\,, \,\,\,i=1,2\quad\text{and}\quad \p=\frac{\partial L}{\partial \dot \r} \in T_{\r}^*\mathbb{R}^3 \simeq \mathbb{R}^3\,.
\end{align}

In order to obtain the reduced Hamiltonian we use the reduced Legendre transform. First we calculate the momenta
\begin{align}
&\Pii_1= \frac{d}{dt}\left( \frac{\d l}{\d \Om_1} \right)= \mathbb{I}_1 {\Om_1}  + m \Gama \times \left( \dot\Gama + \Om_1 \times \Gama \right)  
\label{leg_1}\\
&\Pii_2= \frac{d}{dt}\left( \frac{\d l}{\d \Om_2} \right)=\mathbb{I}_2\, {\Om}_2
\label{leg_2}  \\
&\P=\frac{d}{dt}\left(\frac{\d l}{\d\dot\Gama} \right)=m\left( \dot\Gama + \Om_1 \times \Gama \right)\,.
\label{leg_3}
\end{align}
Next we  calculate the reduced Hamiltonian via 
\begin{align}
H (R, \,&\Pii_1, \Pii_2, \Gama, \P) =\left< \Pii_1, \Om_1(\Pii_1, \Pii_1, \Gama, \P) \right>_{\mathbb{R}^3} + \left< \Pii_2, \Om_2(\Pii_1, \Pii_1, \Gama, \P) \right>_{\mathbb{R}^3}  \nonumber  \\
& + \left< \P, \dot \Gama(\Pii_1, \Pii_1, \Gama, \P) \right>_{\mathbb{R}^3}   -  l \left(\Om_1(R, \Pii_1, \Pii_1, \Gama, \P), \Om_2(\Pii_1, \Pii_1, \Gama, \P) , \Gama, \dot \Gama(\Pii_1, \Pii_1, \Gama, \P) \right)
\end{align}
and obtain the reduced Hamiltonian of the full two body problem  
\begin{align}
&H :   SO(3) \times so^*(3) \times so^*(3) \times T^*\mathbb{R}^3 \to \mathbb{R},\\
&H (R, \Pii_1, \Pii_2, \Gama, \P) 
= \frac{1}{2} \left< \Pii_1+ \Gama \times \P\,, \,\mathbb{I}^{-1}_1 (\Pii_1+ \Gama \times \P)\right>_{\mathbb{R}^3}+\frac{1}{2} \left< \Pii_2, \mathbb{I}^{-1}_2 \Pii_2 \right>_{\mathbb{R}^3}\nonumber \\
&\hspace{9.5cm} +\frac{1}{2} m \left< \P, \P\right>_{\mathbb{R}^3}  
+V(R,\Gama). 
\end{align}
%
%
%
%
%
The dynamics is given by the Poisson bracket 
\begin{align}
 \{F, H\} (R, \Pii_1, \Pii_2, \Gama, \P) &= - \left< \Pii_1,  \frac{\d F}{ \d \Pii_1} \times  \frac{\d H}{ \d \Pii_1}  \right>_{\mathbb{R}^3} -\left< \Pii_2, \frac{\d F}{ \d \Pii_2} \times \frac{\d H}{ \d \Pii_2}\right>_{\mathbb{R}^3}  + \left( \frac{\d F}{ \d \Gama}   \frac{\d H}{ \d \P} -  \frac{\d H}{ \d \Gama}  \frac{\d F}{ \d \P}\right)\\
 &\quad \quad - \left<\frac{\d F}{\d R}\,,\, \frac{\d H}{\d \Pii_1} R - R   \frac{\d H}{\d \Pii_2}\right>
 + \left<\frac{\d H}{\d R}\,,\, \frac{\d F}{\d \Pii_1} R - R   \frac{\d F}{\d \Pii_2}\right>\,.
 \label{Poisson_bra_1}
\end{align}
This is  deduced  by considering the composition of real valued (smooth) functions $F: SO(3) \times so^*(3) \times so^*(3) \times T^*\mathbb{R}^3 \to \mathbb{R}$ with the Poisson map 
\begin{align}
&\lambda: T^*SO(3) \times T^*SO^*(3) \times T^*\mathbb{R}^3\to SO(3) \times so^*(3) \times so^*(3) \times T^* \mathbb{R}^3 \nonumber \\
&\lambda (R_1, \pi_{R_1}, R_2, \pi_{R_1}, \r, \p) = \left( R_1^{-1}R_2, \,R_1^t \pi_{R_1}, \,R_2^t \pi_{R_2}, \r, \p  \right);
\end{align}
using the chain rule,  the canonical bracket on $T^*SO(3) \times T^*SO^*(3) \times T^*\mathbb{R}^3$ becomes the Poisson bracket    \eqref{Poisson_bra_1} (for details on this kind of techniques, see \cite{KM87}).
%
%

%

%
The equations  of the reduced dynamics are:
  \begin{align}
&\dot\Pii_1 = 
\Pii_1 \times
 \left[  \mathbb{I}_1^{-1} (\Pii_1+ \Gama \times \P)  \right] 
  + \sum \limits_{i=1,2,3} \boldsymbol{\alpha}_i \times \frac{\d V}{\d \boldsymbol{\alpha}_i}
    \label{Ham-eq_1}
\\
&\dot\Pii_2 = \Pii_2 \times  \mathbb{I}_2^{-1} \Pii_2
-\boldsymbol{\alpha}_i \times \frac{\d V}{\d \boldsymbol{\alpha}_i}     \label{Ham-eq_2}
\\
&\dot{\Gama}= \frac 1 m \P +\Gama \times \left[  \mathbb{I}_1^{-1}(\Pii_1+  \Gama \times \P)  \right]       \label{Ham-eq_3}   \\
&\dot \P= \P \times  \left[  \mathbb{I}_1^{-1}(\Pii_1+  \Gama \times \P)  \right]   - \frac{\partial V}{ \partial \Gama}
   \label{Ham-eq_4}
\end{align}
together with the reconstruction (orientation) equation:
 \begin{equation}
 \label{R-dot-eq}\dot{R}= R \, \hat \Omega_2 - \hat\Omega_1  \, R.
 \end{equation}
where $R = [\boldsymbol{\alpha}_1\,,\boldsymbol{\alpha}_2\,,\boldsymbol{\alpha}_3]$ and $\hat \Omega_1$ and $\hat \Omega_2$ are calculated via the inverse of \eqref{leg_1}- \eqref{leg_3}.

\bigskip
\begin{remark} Note that with  the choice of ${\cal B}_1$ as reference frame, $\Pii_1$  is the  sum of the angular momentum $\mathbb{I}_1\Om_1$ of the rigid body ${\cal B}_1$ and the angular momentum $\Gama \times \P$ of the relative vector, both  in the body coordinates of ${\cal B}_1$:
\begin{equation}
\Pii_1= \mathbb{I}_1\Om_1+ \Gama \times \P\,.
\end{equation}
\end{remark}

\begin{remark}
 The change of variable \[(\Pii_1, \Pii_2, \P)=(\Lam_1, \Lam _2, \P) := ( \Pii_1- \Gama \times \P, \Pii_2, \P), \] is  a Poisson map (see \cite{Ma92}, Section 3.7) and it leads to the  Hamiltonian  of the two full body problem as used by  \cite{Mac95} and \cite{CM04}:
 \begin{equation} 
 H (R, \Lam_1, \Lam_2, \Gama, \P) = \frac{1}{2} \left< \Lam_1, \mathbb{I}^{-1}_1\Lam_1\right>_{\mathbb{R}^3}+\frac{1}{2}\left< \Lam_2, \mathbb{I}^{-1}_2 \Lam_2 \right>_{\mathbb{R}^3} +\frac{1}{2m}   \P ^2  +V(R,\Gama).
 \end{equation}
 %
The equations of motion are
\begin{align}
&\dot \Lam_1 =  \Lam_1 \times \mathbb{I}_1^{-1} \Lam_1  +\sum \limits_{i=1,2,3} \boldsymbol{\alpha}_i \times \frac{\d V}{\d \boldsymbol{\alpha}_i} + \Gama \times \frac{ \partial V}{ \partial \Gama}\\
&\dot \Lam_2 =  \mathbb{I}_2^{-1} \Lam_2 -  \sum \limits_{i=1,2,3} \boldsymbol{\alpha}_i  \times \frac{\d V}{\d \boldsymbol{\alpha}_i} \\
&\dot \Gama=  \frac{1}{m}\P + \Gama \times \mathbb{I}^{-1} \Lam_1\\
& \dot \P = \P \times \Gama_1 - \frac{\partial V}{\partial \Gama}\,.
\end{align}
Note that this equations coincide to those in \cite{Mac95}.

 \end{remark}

\noindent 
The \textit{spatial total angular momentum} corresponds to the right $SO(3)$ action on the phase space
 it is given  by
  \begin{align} 
 \label{LP-eq}
&J: T^*\left(SO(3) \times SO(3) \times \mathbb{R}^3\setminus \{\text{collisions}\} \right))  \to so^*(3) \nonumber \\
& \hspace{1cm} J\left( R_1, R_2, \r, \breve{\pi}_1,  \breve{\pi}_2, \p \right) \to  (\breve{\pi}_1 R_1^t\ + \breve{\pi}_2 R_2^t + \hat{\r \times \p}  )\,. 
\end{align}  
where we deduced the above using the cotangent bundle momentum map formula (see \cite{HSS09} page 284) and the infinitesimal generator \eqref{inf_so(3)}.
Since the Hamiltonian \eqref{Ham_2_B}  is invariant under the aforementioned action,  by Noether's theorem, the spatial angular momentum is conserved along any trajectory. Denoting $ \Lam _1$ $\Lam _2$  the body angular momenta of ${\cal B}_1$ and ${\cal B}_2$, respectively (i.e., $\Lam_1 =  \mathbb{I}_1 \Om_1$  and $\Lam _2 =  \mathbb{I}_2 \Om_2$) we have 
  \begin{align}  
\| \breve{\pi}_1 R_1^t\ + \breve{\pi}_2 R_2^t + \hat{\r \times \p}   \| &=  
\| \hat \Omega_1^t \mathbb{I}_1 R_1^t\ + \hat \Omega_2^t \mathbb{I}_2 R_2^t + 
\left( R_1 \Gama \times R_1 \P \right) \hat{\,} \,\, \|   \nonumber \\
 &= \| (R_1 \mathbb{I}_1 \hat \Omega_1)^t  + (R_2 \mathbb{I}_2 \hat \Omega_2)^t +  \left(R_1(\Gama \times  \P) \right)  \hat{\,}  \,\, \nonumber \| \\
 &=\| R_1 \mathbb{I}_1 \Om_1 + R_2 \mathbb{I}_2 \Om_2 +   R_1(\Gama \times  \P)   \,\, \nonumber \| \\
 &=\|  \mathbb{I}_1 \Om_1 + (R_1^{-1}R_2)\, \mathbb{I}_2 \Om_2 +  \Gama \times  \P \,\, \nonumber \| \\
 &=\|   \Lam_1 + R \Lam_2 +  \Gama \times  \P \,\,  \|  =\| \Pii_1 + R \Pii_2    \|
\end{align}  
where we used that the relationship between the spatial and body rigid body angular momenta $\breve{\pi}_i = \hat \Omega_i^t \mathbb{I}_i$  (see \cite{HSS09} Section 1.5).
The composition of the spatial momentum  map with the Casimir $C: so^*(3) \to \mathbb{R},$  $C(\x) = || \x ||^2$  leads to the Casimir 
\begin{align} 
&C(R, \Lam_1, \Lam_2, \Gama, \P)=  
\|  \Lam_1 + R \Lam_2 +  \Gama \times  \P   \|^2 = \|  \Pii_1 + R \Pii_2    \|^2
\label{Casimir}
\end{align} 
and  further, any function of the form $\Phi(\|  \Pii_1 + R \Pii_2 \|^2)$ is a Casimir for the reduced dynamics.

\section{Acknowledgements}
The authors  were supported by two Discovery grants awarded by the National Science and Engineering Council of Canada (NSERC).

\section{Appendix}

We append this note with some formulas on the interacting potential. In concordance with most physical situations, we may assume  that  the distance between the bodies is much larger than the bodies dimensions. Thus we consider the potential truncated to the third order  (\cite{Mac95}): 
\begin{align}
V(R, \Gama)= - \frac{Gm_1m_2}{|\Gama |} -  \frac{G}{2|\Gama |^3}\left( m_1 \text{Tr} \,\mathbb{I}_1 + m_1 \text{Tr}\, \mathbb{I}_2  \right) 
+ \frac{3G}{2| \Gama|^5} \left(m_1 \left< \Gama, \mathbb{I}_1  \Gama\right>_{\mathbb{R}^3}+ m_2 \left< R\Gama\,, \mathbb{I}_2R\Gama   \right>_{\mathbb{R}^3} \right) 
\end{align}
where the rotation matrix $R \in SO(3)$ is represented by
 $R= \left[\boldsymbol{\alpha}_1\,, \boldsymbol{\alpha}_2, \boldsymbol{\alpha}_3 \right]$ 
with $\boldsymbol{\alpha}_i$  (column) vectors such that $\boldsymbol{\alpha}_i^2=1$ and $\boldsymbol{\alpha}_i \cdot \boldsymbol{\alpha}_j=0$ for $i \neq j\,.$
Next we calculate the terms $ \boldsymbol{\alpha}_i  \times \partial V/\partial \boldsymbol{\alpha}_i$ and $\Lam \times \partial V/\partial \Lam$  occurring in the equations of motion. Denoting  $\mathbb{I}_2 := \text{diag} (I_{21}, I_{22},I_{23})$,  we obtain:
\begin{align}
\boldsymbol{\alpha}_1 \times &\frac{\partial V}{\partial \boldsymbol{\alpha}_1} = 
2 \Gamma_1^2 
\left(
\begin{array}{c}
 \alpha_2 \alpha_3 (I_{23}-I_{22}) \\
- \alpha_3 \alpha_1 (I_{23}-I_{21})\\
\alpha_1 \alpha_2 (I_{21}-I_{21})
\end{array}
\right) + 2 \Gama_1 \left(   \boldsymbol{\alpha}_1 \times \mathbb{I}_2\boldsymbol{\alpha}_2 + \boldsymbol{\alpha}_1 \times \mathbb{I}_2\boldsymbol{\alpha}_3  \right)
\end{align}
and circular combinations.
Further
\begin{align}
\frac{\partial V} {\partial \Lam } = 
\left[
 \frac{Gm_1m_2}{|\Gama |^2} -
  \frac{3G}{2|\Gama |^4}\left( m_1 \text{Tr} \,\mathbb{I}_1 + m_1 \text{Tr}\, \mathbb{I}_2  \right) 
- \right. & \left.\frac{15G}{2| \Gama|^6} \left(m_1 \left< \Gama, \mathbb{I}_1  \Gama\right>_{\mathbb{R}^3}+ m_2 \left< R\Gama\,, \mathbb{I}_2R\Gama   \right>_{\mathbb{R}^3} \right)  
\right]
\frac{\Lam}{|\Lam|} \nonumber \\
& \hspace{2cm}+\frac{3G}{|\Gama|^4} \left(m_1 \mathbb{I}_1 + m_2 R^t    \mathbb{I}_2 R\right) \frac{\Lam}{|\Lam|}\,
\end{align}
and so 
\begin{align}
\Lam \times \frac{\partial V} {\partial \Lam } = 
\frac{3G}{|\Gama|^5} \left[  m_1 \, \Lam \times \mathbb{I}_1\Lam  + m_2\,  \Lam \times \left( R^t    \mathbb{I}_2 R\Lam  \right) \right]\,.
\end{align}
%

%
\makeatletter
\def\@biblabel#1{}
\makeatother

\end{document}